# Misleading Ourselves: How Disinformation Manipulates Sensemaking

STEPHEN PROCHASKA, Center for an Informed Public, Information School, University of Washington
JULIE VERA, Human Centered Design & Engineering, University of Washington
DOUGLAS LEW TAN, Information School, University of Washington
KATE STARBIRD, Center for an Informed Public, Human Centered Design & Engineering, University of Washington

Informal sensemaking surrounding U.S. election processes has been fraught in recent years, due to the inherent uncertainty of elections, the complexity of election processes in the U.S., and to disinformation. Based on insights from qualitative analysis of election rumors spreading online in 2020 and 2022, we introduce the concept of "manipulated sensemaking" to describe how disinformation functions by disrupting online audiences' ability to make sense of novel, uncertain, or ambiguous information. We describe how at the core of this disruption is the ability for disinformation to shape broad, underlying stories called "deep stories" which determine the frames we use to make sense of this novel information. Additionally, we explain how sensemaking's orientation around plausible explanations over accurate explanations makes it vulnerable to manipulation. Lastly, we demonstrate how disinformed deep stories shape sensemaking not just for a single event, but for many events in the future.

# 1. Introduction

Although seemingly simple in theory, U.S. election processes and procedures are often difficult to understand for non-experts, vary from state to state, and are constantly evolving, due to technological advances, perceived vulnerabilities, and political pressure. Due to the complexity of these processes and the uncertainty of elections, these events provide interesting opportunities to study collective sensemaking. Additionally, sensemaking surrounding recent U.S. elections is occurring within an information environment saturated with disinformation (Benkler et al., 2020; Starbird et al., 2023; Prochaska et al., 2023). Disinformation is defined as false or misleading information, intentionally spread for political, financial, or reputational gain (Starbird et al., 2019). Evidence suggests that disinformation around the U.S. 2020 election — which we conceptualize here as a manipulation of collective sensemaking around election processes and results — played a role in the January 6th, 2021 insurrection at the U.S. Capitol (House of Representatives, 2022; Prochaska et al., 2023). Understanding collective sensemaking in this context is therefore both theoretically interesting and of practical concern.

This workshop paper is based on insights from qualitative research on sensemaking about U.S. elections in both 2020 and 2022. Here, we offer the concept of "manipulated sensemaking" to describe how disinformation disrupts our ability to derive meaning from novel, uncertain, or ambiguous data. We situate our work within existing literature at the intersection of sensemaking, storytelling, and rumoring and integrate insights from our qualitative research to provide a mechanistic explanation for how disinformation disrupts informal sensemaking processes. Most notably, we describe how broad, underlying stories called "deep stories" dictate the frames we use to make sense of novel, uncertain, or ambiguous data and how disinformation interrupts this process. We also describe how enacted sensemaking's focus on plausibility misleads audiences. Lastly, we demonstrate how storytelling in 2020 created a deep story that was used for framing in 2022, supporting our proposal that deep stories shape the interpretive frames we use not just in the present, but in the future.

# 2. Relevant Concepts

## 2.1. Collective Sensemaking

To situate our discussion, we call upon Karl Weick's work, particularly the seven properties of sensemaking as described in *Sensemaking in Organizations*. Weick describes sensemaking as a process that is: grounded in identity construction, social, and driven by plausibility rather than accuracy (Weick, 1995). Weick's assertion that sensemaking is grounded in identity is an important one: people can make sense of experiences if they are rooted in some predictable roles. Roles or identities are typically less volatile than environments and they can be used to anchor sensemaking during rapidly changing conditions. As a social exercise, Weick argues that sensemaking can never be solitary: "what a person does internally is contingent on others" (1995). People make sense of their thoughts and environment within a social context. Shared meanings arise through interactions and negotiations between members of a network. Lastly,

Weick asserts that sensemaking is driven by plausibility rather than accuracy. One of the reasons that people rely on plausibility is that "it is almost impossible to tell, at the time of perception, whether the perceptions will prove accurate or not" (1995). Additionally, people often rely on imprecisely remembered past experiences to elaborate on cues that trigger sensemaking. Weick recognizes that "A good story holds disparate elements together long enough to energize and guide action, plausibly enough to allow people to make retrospective sense" of them, an observation that is highly relevant to the stories of perceived election fraud we examine (1995).

*Data-Frame Theory*
Weick's model of sensemaking has been extended through the Data-Frame Theory of Sensemaking as defined by Klein et al. (2007). This theory postulates that "elements are explained when they are fitted into a structure that links them to other elements" (Klein et al. 2007). Klein et al. use the term *frame* to define the explanatory structures that connect ideas to other explanatory structures. The term synthesizes concepts from decades of work on frames (Goffman, 1974; Minsky, 1975; Rudolph, 2003; Smith et al. 1986) and similar phenomena. Frames can be *stories*, *maps*, *scripts* (Schank & Abelson, 1977), *schemata* (Barlett, 1932; Neisser, 1976; Piaget, 1952, 1954), or *plans*, among other forms. The frame dictates what data we pay attention to, however, they are careful to note neither the data nor the frame necessarily comes first (Klein et al. 2007). Additionally, Klein et al. recognize sensemaking's goal of deriving plausible explanations but observe that "people are explanation machines" and find connections even where there are none (Klein et al. 2007).

*The Fragility of Sensemaking*
Though sensemaking has been characterized as a collaborative endeavor in the progression toward a goal or common understanding, sensemaking processes can sometimes break down. Weick has noted several times and places where sensemaking does not go as planned, including a large wildfire at Mann Gulch that spread out of control. Weick's description of the collapse of sensemaking in the Mann Gulch case study provides a model for how to think about the fragility of sensemaking. Through his analysis, Weick conceptualizes *cosmology episodes,* where sensemaking and underlying structures of rational, orderly conduct simply collapse. "People…act as if events cohere in time and space and that change unfolds in an orderly manner" when this is not necessarily true in reality (Weick 1993). Many events in the Mann Gulch disaster did not form a coherent picture for the people who needed to make sense of the event to successfully put out the erratically-behaving fire. For example, firejumpers thought less of the danger of the fire after learning that other colleagues had stopped their progress to break for dinner. Based on observations from the Mann Gulch case study and others Weick examines, he suggests that there may be a tradeoff between cohesion and accuracy in groups (Weick, 1990, Janis 1982).

Throughout Weick's examples, there is an assumption that people participating in the sensemaking process are genuinely trying to make sense of their environments. Although these breakdowns are invaluable for seeing where good-faith sensemaking breaks down, they don't fully capture what happens when there are members of a sensemaking community intentionally trying to disrupt the process, something our work begins to illuminate.

## 2.2 Rumoring as a byproduct of sensemaking

Rumors are often a byproduct of collective sensemaking. As audiences try to understand the world around them, rumors offer explanations that connect disparate pieces of data. Importantly, rumors can be either true or false, what makes them a rumor is not their facticity but their unofficial source (Kapferer, 1990). According to Kapferer, a rumor consists of "information that is not yet publicly confirmed by official sources or denied by them" (1990). The same uncertainty and ambiguity that lead to sensemaking often give rise to rumors. Although rumors take different forms, they frequently take the form of a story that suggests causal relationships between different story elements. Kapferer describes how rumors often act as exemplary stories: stories that function as examples of larger-scale phenomena believed or suspected by the public (Kapferer, 1990). Exemplary stories seem like stories that should have a moral to tell but are often linked to no tangible facts. Kapferer notes that these stories might not share *the* truth, but they do share *a* truth and have "implications not only for oneself but for the collectivity as a whole" (Kapferer, 1990).

## 2.3 Deep Storytelling with an Online Public

Sociologist Arlie Hochschild's concept of deep stories is central to our work. In "Strangers in Their Own Land" Hochschild identified a broad, underlying story that lay beyond the everyday stories many Americans tell about themselves. This "deep story" was built on people's perceptions of their lived experiences, it was a story that "traded in feelings more than confirmable facts" (Hochschild, 2016). Poletta and Callahan extended the theory of deep stories to the Trump era and explained how the storytelling characteristics of allusiveness, social storytelling, and collective identity construction all played roles in informing the telling of specific stories that informed the underlying deep story (2017). For them, allusiveness is the tendency of stories to have a "moral" that is not explicitly told to audiences, instead audiences must interpret the moral through the events of the story itself. In addition, they explain that storytelling is an inherently social process, and plays a significant role in the creation of collective identity through the implicit identification of norms and values (Poletta & Callahan, 2019).

Social storytelling has been identified as a major component of engagement on social media, with recent work outlining how the intersection of stories of personal experience, social media virality, and political communication can be misleading. Specifically, Mäkelä et al., 2021 describe how the sharing of personal stories online creates a sense that those stories are representative of collective experience that are sufficient to provide a sense of "good cause" for future action/belief, what for Klein et al. might be described as the inference of a cause-effect relationship from the story that is sufficient to act upon.

# 3. High Level Summary of Data and Methods

Our analysis relies upon tweets collected during the 2020 and 2022 elections and later determined to be related to specific rumors curated by our research team. From each election, we selected five salient rumors and used keyword terms and date ranges to identify tweets related to each rumor (see Appendix 1 for the list of rumors examined). During sampling, we ensured that the sample for each rumor included highly spread tweets along with more general conversation. We used a grounded approach to develop a qualitative codebook designed to examine storytelling processes as they relate to the sharing and interpretation of rumors. We coded final samples of 100 tweets per rumor, resulting in a final coded dataset of around 1000 tweets, 500 per year.

---

# 4. Preliminary Results

In the following sections we introduce the concept of "manipulated sensemaking" and how disinformation functions by disrupting our ability to derive meaning, not just our ability to connect disparate pieces of information. After introducing the concept of manipulated sensemaking, we propose three mechanisms that work together to aid the effectiveness of disinformation. First, we describe how deep stories determine the frames we use to make sense of novel, uncertain, or ambiguous data and how disinformation interrupts this process. Second, we discuss how enacted sensemaking's focus on plausible explanations was used to mislead audiences. Lastly, we describe how storytelling in 2020 created a deep story that no longer needed to be described, and was primarily used for interpretation in 2022, supporting our proposal that deep stories shape the interpretive frames we use not just in the present, but for the future.

## 4.1 Overall Finding: Disinformation Manipulates Sensemaking

Disinformation was highly spread during the 2020 election season, consisting of hundreds of false and misleading claims sowing doubt about a variety of aspects of U.S. election processes and procedures, resulting in an overall sense among conservatives that the election was fraudulent (Election Integrity Partnership, 2021; Benkler, 2020). Disinformation in its most basic form is defined as intentionally disseminated false or misleading information. In practice, disinformation is often connected to an overall campaign of deceptive strategies, called "active measures" by Soviet practitioners, whose goal is to mislead audiences as to the reality of a situation (Bittman, 1985; Rid, 2020; Starbird, 2019). Participatory disinformation, the process whereby audiences, both witting and unwitting, collaborate with influencers and political elites to create and spread disinformation, was highly present in the 2020 election (Prochaska et al. 2023, Starbird et al., 2023). In this paper, we describe how disinformation works through the manipulation of collective sensemaking processes.

We propose that the presence of disinformation disrupted normally occurring informal sensemaking processes around elections, taking advantage of sensemaking errors that untrained populations tend to exhibit, particularly a form of frame fixation (Klein et al. 2007). For Klein et al., frame fixation describes the error that occurs when someone trying to understand a situation is stuck on a single explanation (2007). In the case of the data we examined, the disinformation campaign relied strongly on previous rhetoric establishing a (false) frame that the election was rigged and somehow fraudulently stacked against former President Trump. In particular, Trump and his allies continuously, and falsely, suggested that mail-in voting was an insecure method of voting and that Democrats were using this insecurity to commit voter fraud (Benkler et al, 2020).

One of the ways this functioned was by taking advantage of credibility signals the public often use during informal collective sensemaking. As we engage with new information we have no personal experience with, we rely on a collection of heuristics to socially determine whether or not we should trust it and collectively construct our understanding of it. One of the heuristics we rely on is repetition by in-group members (Foster, 2022), and another is the perceived credibility and authority of the source of the information (Lin et al., 2016). In our data we observe both of these heuristics being actively leveraged to influence the construction of a deep story of voter fraud which in turn is used by the audience to make sense of future events using only frames that resonate with the deep story (see also Prochaska et al., 2023).

For example, Maricopa County, Arizona, experienced extensive rumoring surrounding on-the-ground events in both 2020 and 2022. In 2020, rumors revolved around Sharpies bleeding through ballots — which led audiences to incorrectly converge on an interpretation seeing the Sharpie bleeding as an intentional act against conservative voters to prevent their ballots from being counted. (In fact, the ballots were designed to be used with Sharpie pens and the bleed-through did not invalidate votes.)

Similarly, in 2022, errors with tabulation machines caused delays and prevented voters' ballots from being counted on the spot. The recourse in 2022 was to drop ballots in a box, called "Box 3," where the ballots would then be taken to a separate facility to be manually counted. Audiences shared stories in both text and video format of having the tabulation machine reject their ballots, of being told to deposit their ballot in Box 3, and of having to wait in long lines to vote. As these stories were socially spread and collectively interpreted, many members of the online audience (mis)interpreted these events as supporting the deep story that there was widespread voter fraud perpetrated by Democrats, when the reality is that, though there were real issues in the administration of the voting process, those issues were not intentional and had remedies that we part of the normal voting process. In both 2020 and 2022, we see influencers and political elites use their credibility to sow doubt about election processes, and widespread engagement provided a sense that problems were impactful (i.e. enough to change the results of an election).

## 4.2 Explanatory Finding 1: Deep Stories Determine the Frames We Use to Make Sense of Data

Arlie Hochschild's theory of deep stories, as adapted by Poletta and Callahan, provides a compelling framework that complements Weick's and Klein et al.'s models of sensemaking. Building on Klein et al.'s data-frame theory of sensemaking and extending the theory of deep stories, we propose that deep stories provide a potential explanation for how and why people select only a relatively small range of sensemaking frames as they try to make sense of novel data.

This is a more nuanced take on the error of frame fixation. The closest parallel in our data for frame fixation would be a frame of voter fraud, where regardless of the data being examined, audiences and influencers misinterpreted it as evidence of Democrat-led voter fraud. Our data suggests that while this may describe some of the online discussion, it does not fully account for the variety of frames present in the discourse. In particular, elements of the online conversation seemed to center on a collection of frames instead of a single frame. For example, we saw rhetoric across incidents comparing U.S. elections to "third world elections," consistent claims of censorship by liberal media and technology companies (especially Twitter) against conservative audience members, and general suspicion of government and technology (especially voting technology). Critically, these frames did not occur in isolation, they were repeatedly used by members of the online audience to fill in contextual gaps that the frame of voter fraud was insufficient for.

Additionally, in Maricopa County in 2022 we saw the introduction of a frame of voter disenfranchisement by some members of the audience in place of a frame of voter fraud. Importantly, even though the explanatory frame may have changed, much of the underlying attributes remained consistent: Democrats are still the primary villains, the government is still at fault, the media is still censoring conservative viewpoints, etc. This collection of interpretive frames are related, and we argue they are better described together as a deep story than as individual sensemaking frames whose relationship with one another is unclear.

For illustrative purposes, we interpret the "frame" in framing theory to be an analogy referencing a window frame. Extending this analogy, we propose that deep stories may play a structuring role for the frames through which we interpret information. In the analogy, frames are the boundaries of the windows through which we interpret the world. They determine what we see and attend to in a given situation, allowing us to derive structure and meaning from data that otherwise remains ambiguous to us (Goffman, 1974; Entman, 1993). If frames define the windows through which we view the world, then deep stories are the support beams and columns that hold a structure together. They determine the size, height, width, stability, and orientation of the frames we use to make sense of the world. Although there is a certain amount of flexibility in the creation and application of frames, those frames remain rooted to the positionality of the base structure. The foundation of this structure is made up of our core values, values that are contextualized within our experience of the world in the form of deep stories. Importantly, these stories are dynamic, in a sense they are always under construction (similar to Weick's conception of sensemaking as ongoing). As new elements of the story are constructed, storytellers and audiences rely on already existing frames and pre-existing deep stories to provide scaffolding and stability for nascent narratives that may or may not become permanent additions to the structure, introducing new frames in the process.

Whether or not these renovations become permanent appears to be determined, at least in part, by whether they are considered socially acceptable by the community building the structure. An important thing to note is that deep stories and our orientation towards them are socially constructed, something Weick also recognized as a key component of sensemaking (Weick, 1995). Although we each have an individual relationship with them based on our lived experience, they reflect both individual and community values, identities, beliefs, emotions, and, ultimately, frames. Deep stories represent the evolving sum of the relationships between our individual lived experiences and the stories we hear and tell with and about our shared communities. Not all stories told are integrated into a deep story, and not all members of a community may find a story compelling. Only if a narrative becomes widely accepted by many members of a community does it begin to be integrated into a deep story, and even then it may not be permanent. In their discussion of rumoring, Kapferer highlights the fact that many rumors are temporally situated: they are highly compelling only in a certain moment of time under the right conditions, but as people and communities live and change, a rumor, which often take the form of a story, may no longer provide a compelling explanation for the data it was previously used to explain (Kapferer, 2013). Similarly, as communities integrate novel stories into their collective identities, stories that were previously compelling may begin to lose their luster as new stories or shifts in community priorities emerge. We theorize that in order for a deep story to become a permanent or semi-permanent addition, the story must continuously and closely align with the collective and individual values that make up the foundation of the analogic structure we have described. The further from the foundation, the more tenuous the deep story's potential future is.

## Disinformation isn't an information-level attack, it's a "meaning" level attack

This discussion illustrates a critical vulnerability in collective sensemaking processes: if a disinformation campaign can influence the construction of a deep story, it has the potential to mislead the interpretation of innumerable future events. This is because deep stories anchor the selection of frames through which we interpret the world, limiting any interpretation of relevant future data to frames that resonate with a deep story. Because our understanding of reality is highly socially constructed, disinformation need only target the social construction process itself to exert influence across a community. This explains why the problem posed by disinformation is not rooted in facticity: the truth value of a rumor doesn't matter if our *understanding* of the rumor is what is actually being targeted. Importantly, as is highlighted by Thomas Rid, the social construction of reality is positioned as a tactic by those who spread disinformation, not simply an academic theory of knowledge (2020).

## 4.3 Explanatory Finding 2: Stories were Used to Make Plausible Cause-Effect Inferences

One of the pieces Weick and Klein et al. highlight in their discussions of sensemaking is that people work towards plausible explanations instead of perfect explanations of events. The informal sensemaking we observed in our data followed this pattern closely. Stories were shared online with influencers often providing updates on novel events or headlines that audiences in turn added to, often adding detail or including their own personal experience in relation to the item originally shared. As audiences made sense of the collection of stories and interpretations that coalesced into the deep story of voter fraud, functional calls to action emerged in attempts to direct behavior towards solving the perceived problems. For some, the calls were for audits or increased oversight, for others calls included firing those responsible or arresting them if an administrator or candidate was seen as corrupt. In 2020, rallies and protests were organized and attended, and ultimately the #StoptheSteal campaign was pushed by political elites and influencers and picked by the audience, culminating in the January 6$^{th}$ insurrection.

Audiences shared and made sense of a multitude of stories and claims online, not with a goal of abstract understanding, but with a functional goal of understanding if there was a threat to their way of life so that if there was, they could prevent the perceived corruption from continuing. Critically for our discussion, the bar of proof for plausibility is *much* lower than the bar established by accuracy. Usually, this makes sense given our goal of being able to act competently in the world. However, our data demonstrate that when disinformation is a significant factor, relying on plausible explanations during sensemaking becomes a critical vulnerability instead of an efficient shortcut.

## 4.4 Explanatory Finding 3: Deep Stories are Created and then Referenced

As introduced in Section 4.2, deep stories are dynamic and shape future interpretations. In 2020, many more members of the online audience were actively constructing the deep story of voter fraud than did in 2022. They did so by sharing stories of events online and collectively interpreting their significance as related to the deep story of voter fraud, often relating events to their own personal experience. Importantly, political elites such as Donald Trump and his close supporters actively claimed that fraud had occurred and amplified "evidence" provided by audiences and influencers. This engagement provided credibility to the manipulated storytelling that it otherwise would have lacked.

Our data suggest that the deep story of voter fraud continued to wield influence in 2022, but that it only needed to be referenced instead of explicitly stated. Notably, the deep story of fraud was kept alive by political elites and influencers who continuously provided "evidence" for audiences to misinterpret, further reinforcing the deep story. Compared to 2020, the involvement of political elites and influencers was much more subtle in 2022. They made explicit claims of fraud far less often; instead, they relied on audiences to interpret the meaning of the events they shared. This provides further support for our proposal that disinformation that targets the construction of a deep story has the potential to mislead multiple instances of future sensemaking, especially if the disinformation itself masquerades as ongoing sensemaking.

# 5. Discussion and Conclusion

In this work, we have presented a framework that provides a mechanistic explanation for how collective sensemaking is influenced by deep stories. We highlighted how disinformation functions to manipulate informal sensemaking by taking advantage of sensemaking's focus on plausibility, which in turn shapes the deep stories we use to direct our selection of sensemaking frames. Below we describe how our framework increases our understanding of informal collective sensemaking in an adversarial environment as well as how traditional sensemaking practices present vulnerabilities when transferred into a context where disinformation is prevalent.

## 5.1 Mechanism Driving Informal Sensemaking

Our work suggests that the models of sensemaking introduced by Weick and Klein et al. apply to sensemaking surrounding U.S. elections, but that the models do not fully account for an adversarial environment (1993; 2007). By providing a theoretical framework that connects collective sensemaking with both disinformation and deep storytelling, we aim to increase our understanding of how people make sense of uncertain information in contentious environments. Klein et al. examined sensemaking in formal environments, suggesting that expertise comes from having access to higher quantity and quality of frames than amateur sensemakers. Our analysis examines a case study of informal sensemaking (U.S. elections), where many participants are amateur sensemakers (according to Klein et al.'s definition) and primarily do not have expertise surrounding election processes.

Central to our discussion is a recognition that people, particularly in informal situations outside of their expertise, rely on community-informed deep stories to make sense of novel, uncertain, or ambiguous information. An outcome of our analysis is support for the observation that people are not value-agnostic information processing machines: our values, beliefs, and lived experience are encoded in our deep stories which in turn influence the frames we choose to make sense of a situation. Our work suggests that deep stories are particularly salient in informal sensemaking settings. We propose that the manipulated sensemaking framework can increase our understanding of collective sensemaking around U.S. elections, and uncertain, contentious events in general.

## 5.2 Misleading Ourselves: Vulnerabilities in Informal Sensemaking

Simultaneous to improving our understanding of informal sensemaking in adversarial environments, our work highlights specific vulnerabilities in the sensemaking process. Specifically, we describe how people's tendency to focus on plausibility becomes a vulnerability when disinformation is a concern, especially when contextualized within the larger vulnerability presented by the socially constructed nature of deep stories. On their own, both of these mechanisms are adaptive, and as Klein et al. point out, expert sensemakers often perform better specifically *because* they are able to narrow their sensemaking to only likely plausible explanations (Klein et al., 2007). When put into the context of our discussion, it becomes clear that when the context of sensemaking shifts, so too must our sensemaking practices.

In the case of election sensemaking, participatory disinformation was successful because it was a collaborative process by which audiences guided and were guided toward a particular interpretation of a given event. The audiences engaged in enacted sensemaking, but because they did so without full awareness of the strategic nature of the communication they engaged with, they actively participated in misleading themselves and others. Our data demonstrate that this process occurred over time across numerous rumors. The deep story of voter fraud became so salient because the affordances and networked nature of social media facilitated the rapid dissemination and interpretation of "events" that immediately spawned rumors if the event itself wasn't already a rumor. As we describe above, this occurred through repetition over time by audiences, influencers, and political elites. The end result was a deep story that was used for future sensemaking, which significantly reduced the number of sensemaking frames audiences used. This reduction in turn made sensemaking difficult or impossible without re-engaging with frames that were discordant with the audience's newly constructed deep story.

**Acknowledgements**

This material is based upon work supported by the National Science Foundation under Grant Numbers 2120496 and 1749815. Any opinions, findings, and conclusions or recommendations expressed in this material are those of the authors and do not necessarily reflect the views of the National Science Foundation. This work is also supported by funding from the John S. and James L. Knight Foundation, the University of Washington's Center for an Informed Public, and the Election Trust Initiative. The Human Subjects Division determined that the proposed activity does not involve human subjects, as defined by federal and state regulations.

# Work Cited

# Appendix

**Appendix 1:** Table listing the incidents that were sampled and qualitatively coded.

| Incident name | Year | Description |
| --- | --- | --- |
| Konnech Election Software | 2022 | An incident centering around Konnech founder Eugene Yu being arrested on suspicion of personal data theft leading to allegations of election fraud and voting machine security. |
| Lake Amplifies Personal Stories | 2022 | An incident in AZ centering around Kari Lake promoting personal stories of AZ voters experiencing tabulator and printer, voting machine, immigration, and access issues to suggest that these issues are widespread and voting suppression is happening |
| Maricopa Broken Machines | 2022 | An incident in Maricopa County, AZ centering around voting machines not reading ballots leading to it being scanned again or sent "downtown," which fueled claims of fraud. |
| More Votes for Treasurer than Governor | 2022 | In an incident in Arizona, a Republican candidate won the treasurer race, but the Republican candidate for governor lost, sparking allegations of fraud/cheating due to this discrepancy. |
| USPS Dejoy | 2022 | An incident centered around claims that allowing US postmaster Louis Dejoy to continue being in charge would lead to fraud with mail-in ballots. |
| USPS 300,000 Undelivered Ballots | 2020 | An incident centering around claims that the USPS has not delivered 300,000 mail-in ballots on election day, fueling allegations of fraud and obstruction. |

| Ballot Harvesting: Ilhan Omar | 2020 | An incident centering around a video alleging that Minnesota representative Ilhan Omar is utilizing ballot harvesters to exchange cash for absentee ballots. |
| --- | --- | --- |
| Dead Voters | 2020 | An incident in Michigan, allegations that dead people are voting utilizing evidence from plugging in the names of deceased people into a government website. |
| Hammer and Scorecard Conspiracy Theory | 2020 | An incident centering around allegations that Hammar and Scorecard software was being utilized to switch votes in battleground states. |
| Dominion Conspiracy Theory | 2020 | An incident centering around allegations that election software by Dominion Voting Systems had systematically changed votes from Trump to Biden. |